%% file: main.tex
\pdfoutput=1
\documentclass[conference]{IEEEtran}
\IEEEoverridecommandlockouts
\IEEEpubid{\makebox[\columnwidth]{978-1-6654-3540-6/22~\copyright~2022 IEEE \hfill}%
\hspace{\columnsep}\makebox[\columnwidth]{ }}

\usepackage{url}
\usepackage{cite}
\usepackage{amsmath,amssymb,amsfonts}
\usepackage{graphicx}
\usepackage{textcomp}
\usepackage{multirow}
\usepackage{indentfirst}
\usepackage[table,xcdraw]{xcolor}      

\usepackage{listings}                   
\usepackage{enumitem}
\usepackage{comment}
\usepackage[caption=false,font=footnotesize]{subfig}
\usepackage{subcaption}

\usepackage{adjustbox}
\usepackage{booktabs}                   
\usepackage{colortbl}                   
\usepackage{pifont}
\usepackage{makecell}
\usepackage{tabularx}
\usepackage{array}
\usepackage{float}

\def\BibTeX{{\rm B\kern-.05em{\sc i\kern-.025em b}\kern-.08em
    T\kern-.1667em\lower.7ex\hbox{E}\kern-.125emX}}

\begin{document}


\title{
HandPass: A Wi-Fi CSI Palm Authentication Approach for Access Control

\thanks{The authors are supported in part by the grant \#2020/09850-0, \#2022/00741-0, \#2024/23405-0, \#2024/21006-1 by São Paulo Research Foundation (FAPESP), and also from CNPq (312294/2025-5) and CAPES.}
}



\author{
    \IEEEauthorblockN{
        Eduardo Fabricio Gomes Trindade, 
        Felipe Silveira de Almeida, 
        Gioliano de Oliveira Braga\\
        Rafael Pimenta de Mattos Paixão, 
        Pedro Henrique dos Santos Rocha, 
        Lourenco Alves Pereira Jr.
    }
    \IEEEauthorblockA{Division of Computer Science, Aeronautics Institute of Technology (ITA), Brazil\\
    \{trinda05,fsarj81\}@gmail.com, \{giolianobraga,ljr\}@ita.br}
}

\maketitle
\thispagestyle{plain} \pagestyle{plain} 


\begin{abstract}

Wi-Fi Channel State Information (CSI) has been extensively studied for sensing activities. However, its practical application in user authentication still needs to be explored. This study presents a novel approach to biometric authentication using Wi-Fi Channel State Information (CSI) data for palm recognition. The research delves into utilizing a Raspberry Pi encased in a custom-built box with antenna power reduced to 1dBm, which was used to capture CSI data from the right hands of 20 participants (10 men and 10 women). The dataset was normalized using MinMax scaling to ensure uniformity and accuracy. By focusing on biophysical aspects such as hand size, shape, angular spread between fingers, and finger phalanx lengths, among other characteristics, the study explores how these features affect electromagnetic signals, which are then reflected in Wi-Fi CSI, allowing for precise user identification. Five classification algorithms were evaluated, with the Random Forest classifier achieving an average F1-Score of 99.82\% using 10-fold cross-validation. Amplitude and Phase data were used, with each capture session recording approximately 1000 packets per second in five 5-second intervals for each User. This high accuracy highlights the potential of Wi-Fi CSI in developing robust and reliable user authentication systems based on palm biometric data.

\end{abstract}

\begin{IEEEkeywords}
Channel State Information (CSI), machine learning, authentication, palm recognition 
\end{IEEEkeywords}

\section{Introduction}

Over the years, security systems based on recognition have evolved significantly to authenticate users and limit access, mainly to protect sensitive environments and data. However, the rise in malicious cyber threats has questioned the reliability of traditional authentication methods such as passwords, biometrics, and facial recognition. Additionally, the loss of devices like tokens, cards, and QR codes facilitates forgery actions, compromising security. Thus, there is a need to use new technologies and develop different authentication methods that analyze unique biofeatures of users, such as hands size, finger lengths, and finger angle position, thereby increasing convenience, security, and reliability.

Recently, researchers have been investigating using Wi-Fi \textit{Channel State Information (CSI)} data for sensing activities, leveraging the environment's and individuals' characteristics. This technology was initially implemented to adapt the signal to environmental variations, resulting in more efficient and reliable transmission. However, \cite{ZhengYang2020} highlighted that CSI data enables electromagnetic mapping of the environment, facilitating the identification of signal anomalies and user authentication through signals and gestures. Meanwhile, \cite{ChengChen2023} demonstrated that aspects captured in the CSI of Wi-Fi devices could also be extensively explored for authentication purposes. Despite this, approaches like \cite{Shahzad2013}, \cite{Lee2017}, and \cite{Handa2018} required some user action for authentication. Similarly, our study focuses on the intrinsic biophysical characteristics of each User's palm.

Currently, the potential of CSI-based authentication is widely discussed in the literature. The initiative \cite{Zhao2016} demonstrates that analyzing reflections of a person's body makes it possible to recognize his emotional state. In contrast, \cite{Hernandez2023} presented the feasibility of deploying a system to identify a person using Wi-Fi CSI data for access control. Indeed, its practical implementation still needs to be explored, especially using IoT context devices like the Raspberry Pi. Additionally, using small equipment to collect the necessary information for effective authentication represents a significant challenge, particularly considering these devices' extraction, storage, and processing capabilities.

A significant contribution of our approach lies in utilizing Wi-Fi CSI data to capture the palm's subtle and unique biophysical characteristics. This method can offer several advantages, including operating contactless manners that enhance user convenience and hygiene. Additionally, deploying Raspberry Pi devices as low-cost and versatile data collection tools opens up new possibilities for scalable and flexible authentication systems. This approach also leverages existing Wi-Fi infrastructure, making it a cost-effective solution for enhancing security in various settings, from secure facilities to everyday consumer applications. By addressing the limitations of traditional biometric systems and introducing a novel use of Wi-Fi CSI for palm recognition, this research contributes to the advancement of authentication technologies.

Therefore, this manuscript investigates the feasibility of using low-power devices for Wi-Fi CSI data acquisition in this context. It proposes an innovative approach by using Wi-Fi to capture the unique biophysical characteristics of an individual's palm, complemented by machine learning techniques and classification algorithms. By leveraging the ubiquitous nature of Wi-Fi, this proposal offers a non-intrusive and user-friendly alternative to existing biometric systems. To the best of our knowledge, this is the first study to employ Wi-Fi CSI data collected by Raspberry Pi based on palm biophysical characteristics explicitly targeted at authentication for physical access control, increasing technology readiness. Thus, the study brings the following contributions: 

\begin{enumerate}

\item Labeled dataset capturing Wi-Fi CSI data from the right and left palms of 20 users. Ten males and ten females of varying ages.
\item Access Control System Proposal: A proposed access control system based on biophysical characteristics extracted from Wi-Fi data using Raspberry Pi.
\item Performance Evaluation: Performance evaluation of the algorithms \textit{Support Vector Machine (SVM)}, \textit{Random Forest (RF)}, \textit{K-Nearest Neighbors (KNN)}, \textit{Decision Tree (DT)}, and \textit{Naive Bayes (NB)} for user identification.

\end{enumerate}

The remainder of this study is structured as follows: Section 2 discusses previous work, establishing the context for our research. Section 3 describes the adopted methodology, detailing the approaches and tools used. Section 4 presents the experiments conducted, explaining each step of the execution. Section 5 is dedicated to analyzing and discussing the obtained results and exploring their main implications. Finally, Section 6 concludes the study, summarizing the essential findings and proposing future work.

\section{Related Works}\label{sec:related}

Current studies have explored different CSI data techniques, such as activity recognition, gesture detection, gait analysis, location tracking, presence detection, and general sensing. When applied to authentication, these techniques can significantly impact computer science, wireless networks, and IoT security. Additionally, authentication approaches primarily focus on pattern-based methods, mathematical models, and deep learning, each with distinct advantages and challenges.

\subsection{Pattern-Based Authentication}

Pattern-based authentication is achieved by identifying human behaviors through the variation patterns of CSI. In this context, \cite{Shah2017} proposed a two-factor authentication (2FA) system that used CSI data from Wi-Fi networks to verify the physical proximity between devices. However, this authentication relied on associating the individual with their wireless device or a nearby device, requiring additional equipment and thus falling outside the scope of this study.

The research \cite{Wang2016} inspires our work by profiling human movement by transforming CSI data into spectrograms similar to those generated by Doppler radars. Meanwhile, \cite{Guo2017} presented an approach for authenticating human activities based on the beginning and end variance of the Wi-Fi signal. However, both studies used sensors to learn and characterize the activities and explored the 2.4GHz frequency band, which has less granularity for CSI data collection compared to our experiments conducted with Raspberry Pi devices in the 5GHz.

The study \cite{Tan2016} conducted research focused on fine-grained gesture recognition using a single standard Wi-Fi device without requiring the User to wear any sensors. Similarly, the works \cite{Ma2018}, \cite{Tian2018}, and \cite{Ren2019} also utilized channel state information to recognize user gestures. Despite their recognition capabilities, these studies used laptops for CSI data collection, which is different from the actual context of our research.

The notable study \cite{Chen2022} introduced WiFace, a system using Wi-Fi signals to classify facial expressions by detecting subtle CSI changes from facial muscle movements. Conversely, our study focuses on palm biophysical characteristics (hand size, shape, finger spread, and phalanx lengths), providing a less intrusive and more secure biometric authentication method suitable for access control.

\subsection{Model-Based Authentication}

Model-based recognition leverages mathematics or physics to describe and interpret signal variations caused by human behavior. \cite{Zhang2017} were pioneers in constructing models by quantifying the correlation between the dynamics of CSI values, human movement speeds, and specific activities.

The works \cite{Niu2018} and \cite{Zhang2019} are significant in model-based authentication. The first one, for instance, pioneered the construction of models by quantifying the correlation between the dynamics of CSI values, human movement speeds, body parts, and specific activities. Besides, both studies proposed human activity recognition systems using COTS (commercial off-the-shelf) Wi-Fi devices, successfully distinguishing users through diffraction-based detection models. In the same way, \cite{Kong2021} introduced a multi-user authentication system using a model that measures the Time of Arrival (ToA) of signal propagation to create a profile and the Angle of Arrival (AoA) to separate CSI data by the User. However, these approaches suffer from environmental variations, often requiring specific hardware configurations and high-quality CSI data, which can compromise the accuracy of recognition models, especially in access control activities.

The research \cite{Afshar2022} proposed authentication in multi-user environments using normalization to obtain AoAs and clustering to identify which AoAs corresponded to respective users. However, this study was conducted in a simulation environment with fictitious data, disregarding the variations and noise present in a natural environment. Similarly, \cite{Meneghello2023} study on cleaning and processing Wi-Fi's channel frequency response (CFR) Phase to estimate Doppler shifts in a radio monitoring device to distinguish human activities faced difficulties discerning movements that generate nearly identical Doppler effects, even after retraining, reducing confidence in the authentication capability. These limitations highlight the need for further research to address the challenges of model-based authentication in real-world environments.

In the IoT context, \cite{Zhao2021} presented an authentication system based on HMM models and the Fresnel Zone to robustly and efficiently recognize gestures and extract hidden features from CSI data. Meanwhile, \cite{Cheng2021} modeled human activity using the Markov process, employing multiple Gaussian density functions to fit complex activity patterns. However, both HMM and Fresnel Zone-based models can be strongly affected by interferences observed in the IoT context, leading to errors in user authentication.

\begin{table*}[!ht]
\centering
\caption{Comparison between related works}
\label{tab:comparativo}
\normalsize
\resizebox{.85\textwidth}{!}{%
\begin{tabular}{@{}cccccc@{}}
\toprule
\textbf{Study} & \textbf{Approach} & \textbf{Collection Device} & \textbf{\begin{tabular}[c]{@{}c@{}}Frequency used /\\ Bandwidth \end{tabular}} & \textbf{\begin{tabular}[c]{@{}c@{}}CSI Extraction Tool \end{tabular}} & \textbf{\begin{tabular}[c]{@{}c@{}}Application in \\access control \end{tabular}} \\ \midrule
\rowcolor{gray!10} \cite{Shah2017} & Standards & Laptop & 2.4GHz (20MHz) & Not specified & No \\
\cite{Wang2016} & Standards & Laptop & 5GHz (80MHz) & Linux 802.11n CSI Tool & No \\
\rowcolor{gray!10} \cite{Guo2017} & Standards & Laptop & 2.4GHz (20MHz) & Not specified & No \\
\cite{Tan2016} & Standards & Laptop & 2.4GHz (20MHz) & Not Specified & No \\
\rowcolor{gray!10} \cite{Ma2018} & Standards & Laptop & 5GHz (20MHz) & Open RF Linux 802.11n CSI Tool & No \\
\cite{Tian2018} & Standards & Laptop & 2.4GHz (20MHz) and 5GHz (80MHz) & Linux 802.11n CSI Tool & No \\
\rowcolor{gray!10} \cite{Ren2019} & Standards & Mini-ITX & 5.36GHz (40MHx) & Not Specified & No \\
\cite{Chen2022} & Standards & Laptop & 5GHz (80MHz) & Not Specified & No \\
\cite{Zhang2017} & Models & Pair of Antennas & 5.24GHz & Not Specified & No \\
\rowcolor{gray!10} \cite{Niu2018} & Models & Pair of Antennas & 5.24GHz & Not Specified & No \\
\cite{Zhang2019} & Models & Pair of Antennas & 5.24GHz & Linux 802.11n CSI Tool & No \\
\rowcolor{gray!10} \cite{Kong2021} & Models & Laptop & 2.4GHz to 70MHz) and 5GHz (to 200MHz) & Atheros CSI-Tool & No \\
\cite{Afshar2022} & Models & MATLAB & - & - & No \\
\rowcolor{gray!10} \cite{Meneghello2023} & Models & Netgear X4S AC2600 & 5GHz (80MHz) & Nexmon CSI Tool & No \\
\cite{Zhao2021} & Models & Intel NUC & 5GHz (20MHz) & Not Specified & No \\
\rowcolor{gray!10} \cite{Cheng2021} & Models & Lenovo Desktop & 5.32GHz & Linux 802.11n CSI Tool & No \\
\cite{Uddin2019} & Models & Pair of Antennas & 5GHz (80MHz) & Not Specified & No \\
\rowcolor{gray!10} \cite{Yang2023} & Models & Laptop & 5GHz (80MHz) & Not Specified & No \\
\cite{PBV2024} & Models & ESP32 camera module & 5GHz (80MHz) & Not Specified & No \\ 
\rowcolor{gray!10} \cite{Lin2018} & Deep Learning & Laptop & Not Specified & Linux 802.11n CSI tool & No \\
\cite{Zou2018} & Deep Learning & Laptop & 5GHz (80MHz) & Linux 802.11n CSI Tool & No \\
\rowcolor{gray!10} \cite{Chen2019} & Deep Learning & Laptop & Not Specified & Linux 802.11n CSI tool & No \\
\cite{Ding2019} & Deep Learning & Laptop & 5GHz (20MHz) & Linux 802.11n CSI tool & No \\
\rowcolor{gray!10} \cite{Gu2021} & Deep Learning & Mini PC & 2.4GHz (20MHz) and 5GHz (80MHz) & Linux 802.11n CSI tool & No \\
\cite{Gu2022} & Deep Learning & Mini PC & 5GHz (80MHz) & Linux 802.11n CSI tool & No \\
\rowcolor{gray!10} \cite{Mitra2022} & Deep Learning & Intel NUC & 5GHz (80MHz) & Not Specified & No \\
\textbf{Current study} & \textbf{Standards} & \textbf{Raspberry Pi} & \textbf{5GHz (80MHz)} & \textbf{Nexmon CSI Tool} & \textbf{Yes} \\ \bottomrule
\end{tabular}%
}
\end{table*}

AW-TRBAC, presented in \cite{Uddin2019}, dynamically adjusts access permissions based on user behavior and environmental conditions using Wi-Fi signals. While AW-TRBAC is suited for dynamic scenarios, our approach uses static palm positioning for physical access control applications. Still in the movement context, \cite{Yang2023} proposed WiFID, a system identifying drivers by their unique behaviors and physiological traits via Wi-Fi signals. Although WiFID excels in mobile environments, our study employs static palm authentication for non-dynamic access control.

Around the same theme, \cite{Ding2019} proposed HARNN, a Wi-Fi CSI-based human activity recognition approach using deep recurrent neural networks to recognize different human activities. The research significantly contributed to human activity recognition using CSI information. However, the applicability depends on computational complexity, extensive training requirements, environmental sensitivity, critical aspects in the IoT context, and access control.

The proposal \cite{PBV2024} explored IoT and Wi-Fi-based facial recognition for home security, using an ESP32 camera module to enhance security with password and facial recognition. Our study, on the other hand, uses the 5GHz frequency band with 80 MHz bandwidth, increasing CSI data granularity and utilizing Amplitude and Phase data for user authentication.

\subsection{Deep Learning-Based Authentication}

Deep learning can automatically learn and extract significant features from input data, eliminating the need for manual feature extraction steps. In this regard, \cite{Yousefi2017} compiled essential deep learning techniques, such as recurrent neural networks (RNN) and long short-term memory (LSTM), demonstrating their enhanced performance.

Building on this, \cite{Lin2018} proposed automatic activity and gait segmentation using an automatic segmentation algorithm (ASA), followed by ResNet, to validate legitimate users and recognize unauthorized users. \cite{Zou2018} introduced the Autoencoder Long-term Recurrent Convolutional Network (AE-LRCN) to clean noise from raw CSI data, extract high-level representative features, and reveal temporal dependencies between data for precise human activity recognition. In the same regard, an attention-based bidirectional long short-term memory (ABLSTM) for passive human activity recognition using Wi-Fi CSI signals was presented in \cite{Chen2019}. However, in the context of access control, these approaches face challenges related to computational complexity, the need for large volumes of data for practical training, sensitivity to variations in the physical layout of the environment, and interference from other devices, compromising the robustness of the system in real scenarios.

Deep learning to the physical behavior of users captured by Wi-Fi channel state information (CSI) was applied in \cite{Gu2021} to distinguish legitimate users from impostors. Similarly, \cite{Gu2022} sought to authenticate users by the dynamics of keystrokes during login attempts. However, these studies primarily focused on user behavior and typing rhythm, leaving gaps for attempts to impersonate the physical behavior of authorized users, making these approaches unsuitable for access control applications.

A facial authentication-based digital ID system for smart cities, iFace 1.1, was developed in \cite{Mitra2022} and focused on deepfake and presentation attack detection. In contrast, our study uses Wi-Fi CSI for palm-based biometric recognition, providing a robust and non-intrusive identification method.

Our study differentiates itself by the simplicity of the proposal, using Raspberry Pi for data collection at the 5GHz frequency and employing hand features, all to physical control access through Wi-Fi authentication. 

Table \ref{tab:comparativo} compares related manuscripts with the current study and highlights that there is still room for implementing an access control system using CSI data. In that regard, to our knowledge, this is the first work to use combined features extracted from Wi-Fi CSI by low-power devices with an access control proposal, presenting itself as a complementary security.

\section{CSI}\label{sec:proposal}

This section provides an overview of the properties of Channel State Information (CSI) and the basic principles of user authentication, as well as describes the architecture of the proposed system for access control.

\subsection{CSI Proprieties}

Wi-Fi Channel State Information (CSI) records details about how wireless signals travel from the transmitter to the receiver, describing the behavior of electromagnetic waves across frequencies. This information includes the signal's Amplitude and Phase, which can alter reflections, obstructions, and other environmental factors. Each element of the CSI indicates how the environment impacts signal propagation, forming a function known as the Channel Frequency Response (CFR). The technology divides the spectrum into several subcarriers in Wi-Fi systems using multiple antennas through Orthogonal Frequency Division Multiplexing (OFDM). The transmitter sends unique training signals at the beginning of the transmission, which the receiver uses to estimate how the Wi-Fi channel affects each subcarrier. This helps the receiver understand the signal's behavior under different conditions, allowing adjustments to ensure better data transmission and reception.

\begin{figure}[!b]
    \centering
    \includegraphics[width=\columnwidth]{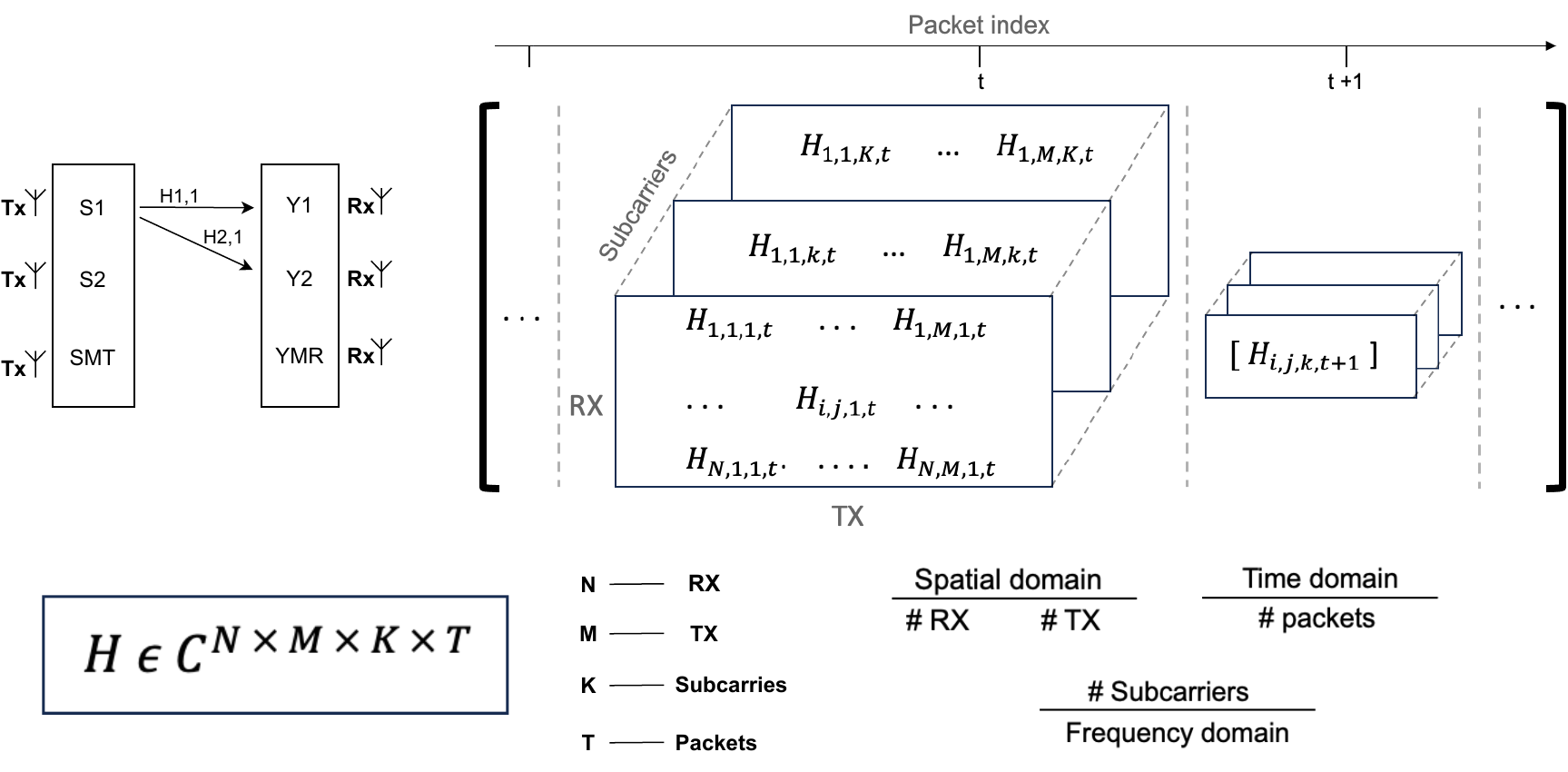}
    \caption{CSI Matrix adapted from \cite{YMa2019}.}
    \label{fig:matrix}
\end{figure}

According to \cite{YMa2019}, the CSI matrix, illustrated in Figure \ref{fig:matrix}, is a three-dimensional set of complex values the receiver estimates from the received signal. This process involves removing the cyclic prefix, demapping, and OFDM demodulation. In practice, the measured CSI is affected by multipath channels, processing at the transmitter and receiver, and inconsistencies in hardware and software. The representation of CSI in the baseband domain is a complex abstraction that considers all these factors, including cyclic shifts and variations in time and frequency sampling.

The time series of CSI matrices captures changes in the MIMO channel over time, frequency, and space. For a MIMO-OFDM channel with M transmitting antennas, N receiving antennas, and K subcarriers, the CSI matrix forms a data cube, expressed as $H \in C^{(N \times M \times K \times T)}$. This cube records how signals undergo Amplitude attenuation and Phase shift as they traverse multiple paths. Thus, CSI provides much richer information than other metrics, such as the Received Signal Strength Indicator (RSSI).

Matrix calculations on raw CSI data result in complex numbers that capture how the signal varies over time. These complex numbers are defined as $z = a + bi$, where $a$ is the real part, $b$ is the imaginary part, and $i$ is the imaginary unit, with the property $i^2 = -1$. Thus, we can obtain the Amplitude (or modulus) of a complex number represented by  $|z| = \sqrt{a^2 + b^2}$. Here, $a$ is the real part, and $b$ is the imaginary part. It is also possible to calculate the PhasePhaseangle) $\theta$, which is the Angle formed by the vector representing the complex number in the complex plane relative to the real axis. The Angle is defined as $\theta = \text{atan2}(b, a)$, where $\text{atan2}(b, a)$ is the arctangent function of two variables that returns the Angle whose tangent is $b/a$, considering the sign of both to determine the correct quadrant. In this way, variations in the Amplitude and Phase of the signal provide a more detailed understanding of the analyzed behavior.

According to \cite{DanWu2017}, identifying patterns in the signal and relating them to specific user characteristics or behaviors makes it possible, through a machine learning-based approach, to accurately identify a user. In this regard, our study encompasses signal preprocessing methods, feature extraction, and the application of machine learning algorithms to classify patterns for physical access control.

\subsection{Authentication with CSI}

According to \cite{Meneghello2023}, using CSI for human authentication explores the idea that different users generate unique CSI variations within the signal's coverage. This means that it is possible to recognize a user's identity by analyzing the dynamic profile of the CSI. However, user authentication using CSI data for physical access control is innovative, especially when the privacy policy is a constant concern.

Essentially, there are two main types of user authentication: one based on CSI fluctuations caused by user movements, such as steps, activities, and gestures, and another that uses the static propagation characteristics of the CSI when the User is stationary. A supervised investigation, labeling the input data, and treating recognition problems as classification facilitates the identification of patterns and regularities in the data.

The action-based approach is widely employed due to the human movement's ability to generate evident fluctuations in the CSI, which can be measured and processed by various available algorithms. On the other hand, immobility-based authentication requires extracting unique biophysical characteristics, such as silhouette, water rate, fat, and muscle, or combining the User's location. Thus, user authentication based on CSI offers a promising way to accurately identify individuals by exploring the nuances of signal variations.

\begin{figure*}[h!]
    \centering
    \includegraphics[width=.75\linewidth]{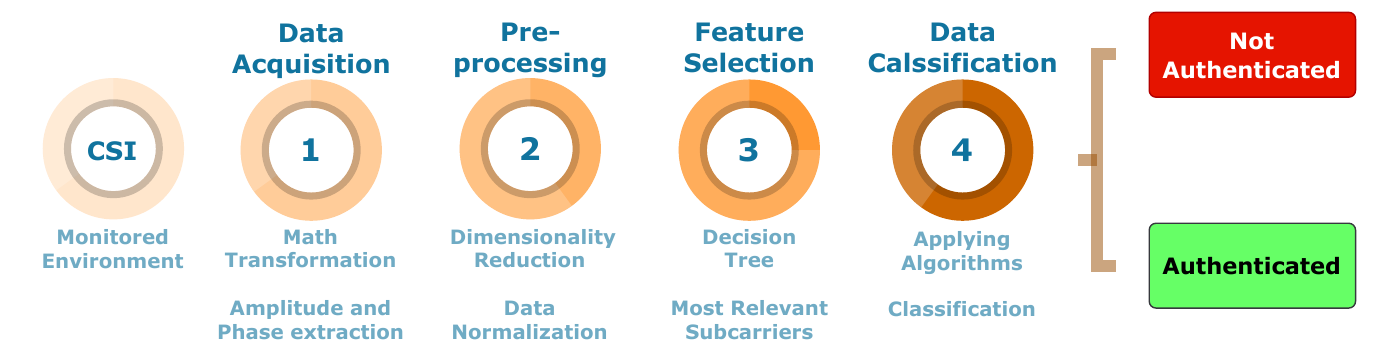}
    \caption{Suggested model.}
    \label{fig:proposal}
\end{figure*}

\subsection{Proposed Access Control System}

This study proposes an authentication system for physical access control in monitored environments, as illustrated in the proposal flowchart in Figure \ref{fig:proposal}. One Raspberry Pi operating in monitor mode initially collects CSI data. During the capture, a transformation from the time domain to the frequency domain is performed, delivering the result in hexadecimal format. Next, the data must be reordered with FFT Shift to center the zero frequency, allowing for conversion into complex numbers. This enables the extraction of the signal's Amplitude and Phase preprocessing.



The data preprocessing involves various operations to extract the desired features. Initially, the CFR values are normalized by the average Amplitude of the 256 monitored subchannels to remove unwanted amplifications. Then, a Phase sanitization algorithm is applied, with an adjustment parameter fixed at $\lambda = 10^{-1}$. The CFR values are then reconstructed and combined, resulting in a complex CFR vector with 512 components containing Amplitudes and Phases.

Normalization is a crucial step in data preprocessing to ensure that features contribute equally to the analysis. This study also compared three normalization techniques: MinMax, Z-Score, and RobustScaler. Before normalization, initially, in radians, the Phase data was converted to degrees to facilitate more intuitive interpretation and processing.

MinMax Scaling scales the data to a fixed range, typically 0 to 1. It is straightforward and widely used due to its simplicity and efficiency. MinMax scaling is particularly effective when the data distribution is not Gaussian and helps mitigate the impact of outliers. Given the uniform range, Amplitude and Phase variations in the CSI data are normalized effectively, facilitating better performance in machine learning algorithms. Our choice for MinMax was also influenced by its slightly better performance in our classification tasks than Z-Score and RobustScaler.

Z-Score Normalization, or standardization, transforms the data into a mean of zero and a standard deviation of one. Z-score normalization is beneficial when the data follows a Gaussian distribution. It handles outliers more effectively than MinMax scaling. However, in our dataset, significant variations in both Amplitude and Phase data made this method less optimal.

RobustScaler scales the data according to the interquartile range, making it robust to outliers. RobustScaler is useful in datasets with many outliers or non-Gaussian distributions. Although it provided reasonable results, the computational complexity and the need for additional processing steps made it less desirable for our application.
After careful consideration, MinMax scaling was chosen for our dataset for its simplicity, efficiency, and slightly better performance in our classification tasks. This approach ensured that the normalized data maintained its integrity, allowing for accurate and reliable feature extraction.

After preprocessing, we analyze how the Amplitudes and Phases vary in each subcarrier according to the captured data, and the \textit{Decision Tree} algorithm is applied to extract the most relevant subcarriers. Next, we develop a machine learning model, trained with 10-fold cross-validation, to estimate each hand user's characteristics. Finally, user authentication is performed through data classification using the \textit{Random Forest} algorithm. This way, authorized users have their hands recognized, and access is granted according to the permissions. Additionally, we compared the model's ability to authenticate users at different times and amounts of data within a security margin considered in this study as acceptable for authentication.

\section{Experiments}\label{sec:experiments}

This section presents how the experiments were conducted, the data collection process, and the capture protocol used. This ensures that the work can be replicated and improved upon in the future.

\begin{figure}[h!]
    \centering
\subfloat[User's hand features capture.\label{fig:capmao}]{
    \includegraphics[width=0.45\columnwidth]{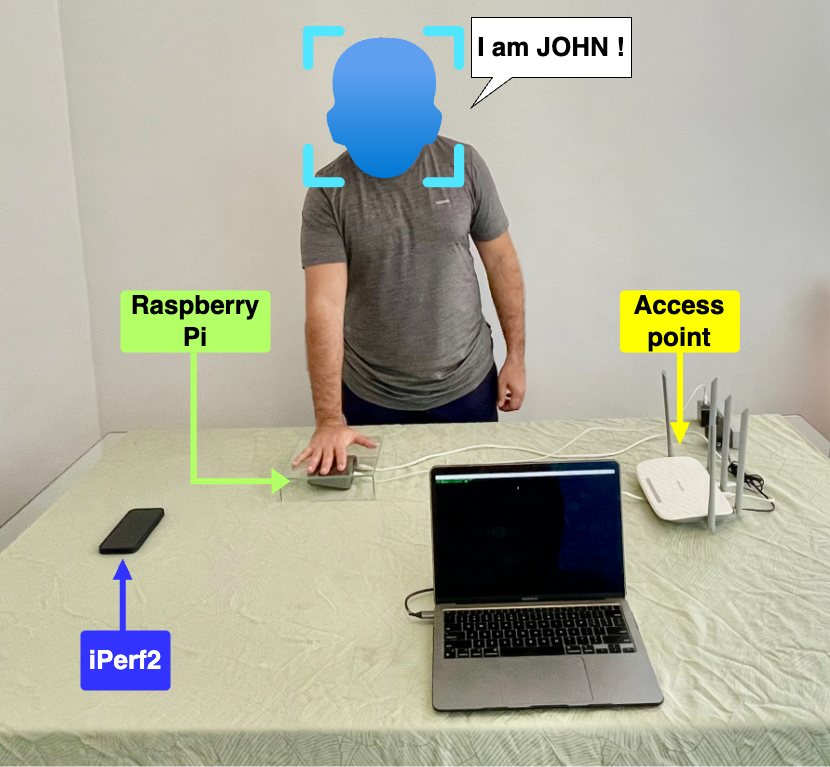}
}
\hfill
\subfloat[Acrylic Box with Raspberry Pi.\label{fig:box}]{
    \includegraphics[width=0.45\columnwidth]{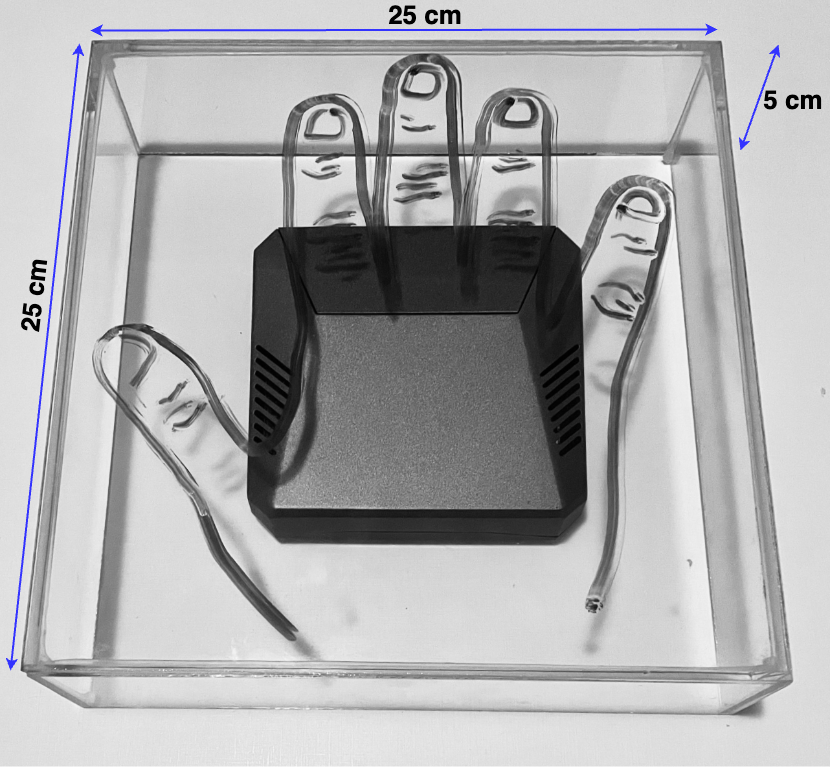}
}
\caption{Experiments.}
\label{fig:capturas}
\end{figure}

\begin{table*}[t!]
\centering
\caption{CSI data collection protocol}
\label{tab:protocol}
\resizebox{.85\textwidth}{!}{%
\begin{tabular}{cc}
\toprule
\textbf{Monitoring User Characteristics} & 20 users (10 men and 10 women) \\ 
\midrule
\textbf{Height - Weight - Age} & 1,65m a 1,85m - 60kg a 93kg - 18 a 63 years old \\ 
\midrule
\multirow{3}{*}{\textbf{Traffic Generation}} & Mobile device with application iPerf2 \\ & Sending 1000 UDP packets per second \\ & Parameters: \texttt{-c 192.168.1.1 -u -b 500M -t 60 -i 1 -l 1400} \\
\midrule
\multirow{2}{*}{\textbf{Hand Positioning}} & 5 Distinct hand positions above the acrylic box \\ & 3cm distance from Raspberry Pi with 1dB antenna \\
\midrule
\multirow{2}{*}{\textbf{Data Capture}} & 100 Instances of 5 seconds each \\ & Recording reflections, attenuations and diffractions of the electromagnetic signal \\
\midrule
\multirow{3}{*}{\textbf{Calibration and Capture}} & 3 second initial delay  \\ & 5 second capture sessions \\ & Line-of-Sight (LOS) mode to avoid obstacles  \\
\bottomrule 
\end{tabular}%
}
\end{table*}

The CSI data was captured in a controlled environment using an acrylic box with a Raspberry Pi to simulate Wi-Fi-based authentication. Each user positioned one hand at a time above the acrylic box for five 5-second intervals. The antenna power of the Raspberry Pi was reduced to 1dBm to minimize potential interference. The CSI data was extracted from the .pcap files and converted into .csv format, with amplitude (in magnitude) and phase (in degrees) information extracted from the 256 subcarriers. These amplitude and phase values were then normalized and combined into CSV files with 512 columns (256 for amplitude and 256 for phase). Additional metadata was appended to the end of the CSV files, resulting in four extra columns: capture number (1 to 5), gender, hand (right or left), and user ID (01 to 20). Thus, each capture file initially contained 516 columns. After removing null and pilot subcarriers, the files were reduced to 472 columns.

From these files, the data from the right hand of each user was isolated, creating a total of 100 files (5 captures per right hand for each of the 20 users). These 100 files were then used to create six distinct datasets, each containing different segments of the captured data to determine the minimum dataset size required for accurate user identification. The first dataset included 1 second of capture number 1 from each user, amounting to 107MB. The second dataset comprised 1 second from captures 1, 2, and 3 from each user, resulting in 323MB. The third dataset contained 1 second from captures 1, 2, 3, 4, and 5 from each user, totaling 536MB. The fourth dataset included 5 seconds of capture number 1 from each user, reaching 648MB. The fifth dataset consisted of 5 seconds from captures 1, 2, and 3 from each user, summing up to 1.58GB. Finally, the sixth dataset encompassed 5 seconds from captures 1, 2, 3, 4, and 5 from each user, resulting in 2.62GB.

All five classification algorithms (Random Forest, Support Vector Machine, K-Nearest Neighbors, Decision Tree, and Naive Bayes) were evaluated across all six datasets to determine the smallest dataset size that could still achieve high accuracy in user identification.

The data capture was conducted in a environment measuring 2 m (width) x 2 m (depth) x 3 m (height), as illustrated in Figure \ref{fig:capmao}. An acrylic box measuring 25 cm (width) x 25 cm (depth) x 5 cm (height) with a Raspberry Pi inside, shown in Figure \ref{fig:box}, was used to simulate Wi-Fi-based authentication to capture the signal anomalies caused by the users' hands. The acrylic box was chosen due to its minimum radio wave absorption and interference capacity, favoring the performance of MIMO and OFDM technologies in Wi-Fi and a more precise CSI data capture.

In this scenario, one Raspberry Pi 4 Model B device was used as a receiver (Rx), equipped with a 64-bit quad-core Cortex-A72 processor, 8GB LPDDR4 RAM, wireless connection capability in the 802.11b/g/n/ac standard, Bluetooth 5.0, PoE capability, and consuming only 5V/3A, which allows them to be powered by a USB-C connected power bank. This equipment was chosen for being undersized (25x52x10mm), with low energy consumption compared to devices used in other studies, and for being ideal for developing numerous automation applications directed towards IoT and machine-to-machine (M2M) communication.

The Raspberry Pi device was remotely controlled by a DELL Inspiron 15 Gaming 7567 laptop with Windows 10 Home, an Intel octa-core i7-7700HQ 2.80GHz processor, and 16GB of RAM, accessed via SSH. To simulate the Wi-Fi signal of a fictitious network, a TP-Link Archer C60 router was used as the transmitter (TX), operating a 5GHz network at 80MHz on channel 36. The capture protocol followed the one in Table \ref{tab:protocol}.

Twenty users with different physical characteristics were monitored, including ten men and ten women. For accuracy and future reproducibility, the users' heights ranged from 1.65 m to 1.85 m, their weights ranged from 60 to 93 kilograms, and their ages ranged from 18 to 63 years old. Additionally, a mobile device was used to generate traffic using the \textit{iperf2} application with a rate of approximately 1000 UDP packets per second. This was executed on an Android system and configured with the following parameters: \texttt{-c 192.168.1.1 -u -b 500M -t 60 -i 1 -l 1400}.

\begin{figure*}[htb]
    \centering
    \includegraphics[width=\linewidth]{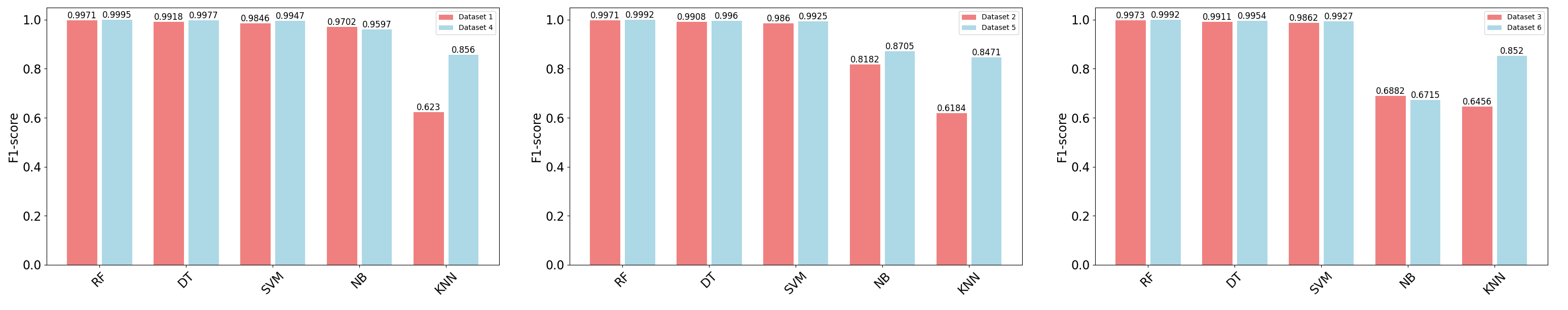}
    \caption{F1-Score performance between datasets.}
    \label{fig:ex1}
\end{figure*}

During the data collection, users positioned one hand at a time above the acrylic box five times, for 5 seconds each time, maintaining a distance of 3 cm from the Raspberry Pi. This distance was limited and standardized by the acrylic box. The device's antenna power was reduced from 31dBm to 1dBm to minimize potential interference from unwanted events or objects. Additionally, users were instructed to remove any watches, bracelets, and rings to reduce interferences. The dataset was comprehensive, comprising 200 captures of 5 seconds each (5 captures of each hand from each User), totaling approximately 1,100,000 instances (5500 per capture) that recorded the reflections, attenuations, and diffractions experienced by the electromagnetic signal along its path in the analyzed environment. However, for this work, we used half of the dataset, i.e., only the data from the right hand.

Additionally, each collection activity had an initial delay of three seconds before the start of recording, and the tool was calibrated so that each capture session lasted five seconds, ensuring the consistency, accuracy, and standardization of the collected data. It is also worth noting that the Line-of-Sight (LOS) mode was used, avoiding obstacles between the transmission and reception points. This approach enhances the robustness of the communication, as it minimizes signal attenuation and dispersion, providing cleaner and more robust communication.

\section{Results}\label{sec:results}

The CSI data collected in the experiments were performed using the Nexmon tool, proposed by \cite{Gringoli2019}. The extracted data included information from 256 subcarriers, with each file containing approximately 5,500 packets. The CSI data is embedded within the payload of the UDP packet incorporated in the PCAP files generated by the tool. The payload content was extracted and converted to CSV files to facilitate data manipulation and analysis. Signal interferences were described by attributes such as Magic Bytes, RSSI, FrameControl Byte, Source Mac, Sequence Number, Core and Spatial Stream, Chanspec, Chip Version, and CSI Data. However, since our study focuses on the analysis of CSI, only the CSI data attribute was considered relevant for the authentication-focused analysis for access control, with the others being discarded.

The CSI Data attribute provided complex numbers from matrix calculations applied to the raw CSI data. After the mathematical transformation, information from the 256 subcarriers was accessed, although some were null (-128, -127, -126, -125, -124, -123, -1, 0, 1, 123, 124, 125, 126, 127) and others were pilots (-103, -75, -39, -11, 11, 39, 75, 103). Therefore, 234 useful subcarriers remained for analysis in each capture.

Three scenarios with different CSI data collection times were tested to obtain a clearer view of the authentication model's capacity. The model was evaluated by attempting to recognize the right hand of each User using one, three, and five-second samples of captured CSI data as shown in Figure \ref{fig:ex1}. The recognition capacity increases with longer capture times, indicating that more data packets result in better authentication performance.

Besides, all algorithms were trained with 10-fold cross-validation, avoiding overfitting and providing a more realistic estimate. The observed differences suggest that it is possible to authenticate the User by their intrinsic characteristics, enabling this information for authentication and access control. It is noteworthy that the CSI is directly influenced by the monitored User, with the correlation of the subcarriers varying over time according to their biophysical characteristics.

\subsection{Model's performance}

The overall average F1-Score was measured using RF, DT, SVM, NB, and KNN algorithms, as shown in Table III. The Comparison of the results shows that larger datasets can slightly improve performance. However, it is known that the authentication process must be easy, fast, reliable, and with reduced friction time, considering the inherence factor. Therefore, this study evaluated the amount of CSI data captured and the capture time to find the algorithm with the best performance. Among these, Random Forest presented the best average F1-Score among all datasets, with 99.82\% in distinguishing users' palms, while the other classifiers achieved 99.38\%, 98.95\%, 82.97\%, and 74.03\%, respectively.

\begin{table*}[t]
\centering
\caption{Average result of classifiers}
\label{tab:classifiers}
\resizebox{.70\textwidth}{!}{%
\begin{tabular}{cccccc}
\toprule
\textbf{Metrics} & \textbf{RF} & \textbf{DT} & \textbf{SVM} & \textbf{NB} & \textbf{KNN} \\
\midrule
\rowcolor{gray!10} \textbf{F1-Score}     & 0.9982  & 0.9938  & 0.9895 & 0.8598  & 0.7103   \\
\textbf{Accuracy}                                    & 0,9982 & 0,9938 & 0,9898 & 0,8630 & 0,7099   \\
\rowcolor{gray!10} \textbf{Precision}    & 0.9982   & 0.9938  & 0.9898 & 0.8850  & 0.7307    \\
\textbf{Recall}                                       & 0.9982   & 0.9938   & 0.9892   & 0.8635   & 0.7107    \\
\bottomrule
\end{tabular}
}
\end{table*}

\textit{Random Forest} was chosen for several reasons, including its higher accuracy and robust performance across various scenarios. The result is reflected by an average accuracy of 0.9981 for recognizing individuals' hand characteristics, indicating a high authentication capability of the model. Its ensemble nature is decisive in our authentication approach, allowing the proposed model to be adaptive and accurate. The classifier makes predictions by aggregating the results of multiple decision trees, which helps capture the nuances in the data and reduces the risk of overfitting.

Furthermore, Random Forest's ability to handle many features and its robustness against noise make it particularly suitable for our application, where CSI data exhibit significant variability. The model's capability to assess the importance of each feature also facilitates understanding which aspects of the CSI data contribute most to accurate authentication, thereby enhancing the model's precision.

Additionally, Random Forest's inherent parallelism allows for efficient training and prediction, making it scalable for large datasets. This is essential for dynamic environments where new data is continuously generated. The algorithm can integrate new users and adapt to changes in the training data without significant reconfiguration, ensuring that the access control system remains robust and reliable over time.



The metrics obtained in the experiments yielded results that enable users to authenticate themselves, allowing the identification of those with authorized access. Furthermore, better performance could be achieved by using more capture equipment, eliminating environmental noise, and increasing the reflective ability of the electromagnetic signal in a fully reflexive box. However, the study was oriented toward a practical application with the least possible equipment.

Our study significantly elevates technological readiness in biometric authentication using Wi-Fi CSI data. The high accuracy achieved with the Random Forest classifier demonstrates the robustness of the proposed system. Beyond access control, this approach has promising applications in various domains requiring secure hand-based authentication, such as financial transactions, secure document handling, and personalized smart home environments. The adaptability and precision of our model suggest a wide range of future applications where reliable and non-intrusive user authentication is critical.

\section{Conclusion and Future Works}\label{sec:conclusion}

This study highlights the use of Wi-Fi data as an innovative method for user authentication in physical access control systems, presenting a significant advancement in biometric authentication by leveraging Wi-Fi CSI data for palm recognition. By employing Raspberry Pi devices in a controlled environment and applying supervised learning techniques, it was possible to identify users with an accuracy of 99.92\% using the Random Forest (RF) classifier. The research validated that biophysical characteristics captured by CSI can be reliably used for user authentication. This offers a secure, non-intrusive, and efficient alternative to traditional access control methods, allowing for confirming credentials without direct interaction by collecting data. The robustness and high accuracy of the Random Forest classifier, due to its ensemble nature and ability to handle large feature sets and noisy data, make it particularly well-suited for dynamic and constantly evolving environments.

Our research provides a robust solution for user authentication and paves the way for future work in this field. We envision expanding the database, integrating more Raspberry Pi devices to capture a more significant number of biophysical characteristics, focusing on optimizing the system for dynamic environments, exploring the integration with other biometric modalities to enhance security further, and studying the feasibility of using unsupervised learning. These potential avenues for further research underscore the significance of our study and its potential to inspire future innovations in user authentication.

\section*{Acknowledgments}

Omitted for double-blind review


\bibliographystyle{IEEEtran}
\input{main.bbl}

\end{document}

%% file: main.bbl

%% file: main.bbl
\begin{thebibliography}{10}
\providecommand{\url}[1]{#1}
\csname url@samestyle\endcsname
\providecommand{\newblock}{\relax}
\providecommand{\bibinfo}[2]{#2}
\providecommand{\BIBentrySTDinterwordspacing}{\spaceskip=0pt\relax}
\providecommand{\BIBentryALTinterwordstretchfactor}{4}
\providecommand{\BIBentryALTinterwordspacing}{\spaceskip=\fontdimen2\font plus
\BIBentryALTinterwordstretchfactor\fontdimen3\font minus \fontdimen4\font\relax}
\providecommand{\BIBforeignlanguage}[2]{{%
\expandafter\ifx\csname l@#1\endcsname\relax
\typeout{** WARNING: IEEEtran.bst: No hyphenation pattern has been}%
\typeout{** loaded for the language `#1'. Using the pattern for}%
\typeout{** the default language instead.}%
\else
\language=\csname l@#1\endcsname
\fi
#2}}
\providecommand{\BIBdecl}{\relax}
\BIBdecl

\bibitem{ZhengYang2020}
Y.~Zhang, Y.~Zheng, G.~Zhang, K.~Qian, C.~Qian, and Z.~Yang, ``Gaitid: Robust wi-fi based gait recognition,'' in \emph{Wireless Algorithms, Systems, and Applications: 15th International Conference, WASA 2020, Qingdao, China, September 13--15, 2020, Proceedings, Part I 15}.\hskip 1em plus 0.5em minus 0.4em\relax Springer, 2020, pp. 730--742.

\bibitem{ChengChen2023}
C.~Chen, H.~Song, Q.~Li, F.~Meneghello, F.~Restuccia, and C.~Cordeiro, ``Wi-fi sensing based on ieee 802.11bf,'' \emph{IEEE Communications Magazine}, vol.~61, no.~1, pp. 121--127, 2023.

\bibitem{Shahzad2013}
\BIBentryALTinterwordspacing
M.~Shahzad, A.~X. Liu, and A.~Samuel, ``Secure unlocking of mobile touch screen devices by simple gestures: You can see it but you can not do it,'' in \emph{Proceedings of the 19th Annual International Conference on Mobile Computing and Networking}.\hskip 1em plus 0.5em minus 0.4em\relax New York, NY, USA: Association for Computing Machinery, 2013, pp. 39--50. [Online]. Available: \url{https://doi.org/10.1145/2500423.2500434}
\BIBentrySTDinterwordspacing

\bibitem{Lee2017}
W.-H. Lee and R.~B. Lee, ``Sensor-based implicit authentication of smartphone users,'' in \emph{2017 47th Annual IEEE/IFIP International Conference on Dependable Systems and Networks (DSN)}, Denver, CO, USA, 2017, pp. 309--320.

\bibitem{Handa2018}
J.~Handa, A.~Singh, A.~Goyal, and P.~Aggarwal, ``Behavioral biometrics for continuous authentication,'' in \emph{2018 Fifth International Conference on Parallel, Distributed and Grid Computing (PDGC)}, 2018, pp. 284--289.

\bibitem{Zhao2016}
\BIBentryALTinterwordspacing
M.~Zhao, F.~Adib, and D.~Katabi, ``Emotion recognition using wireless signals,'' in \emph{Proceedings of the 22nd Annual International Conference on Mobile Computing and Networking}.\hskip 1em plus 0.5em minus 0.4em\relax New York, NY, USA: Association for Computing Machinery, 2016, pp. 95--108. [Online]. Available: \url{https://doi.org/10.1145/2973750.2973762}
\BIBentrySTDinterwordspacing

\bibitem{Hernandez2023}
S.~M. Hernandez and E.~Bulut, ``Wifi sensing on the edge: Signal processing techniques and challenges for real-world systems,'' \emph{IEEE Communications Surveys \& Tutorials}, vol.~25, no.~1, pp. 46--76, 2023.

\bibitem{Shah2017}
\BIBentryALTinterwordspacing
S.~W. Shah and S.~S. Kanhere, ``Wi-auth: Wifi based second factor user authentication,'' in \emph{Proceedings of the 14th EAI International Conference on Mobile and Ubiquitous Systems: Computing, Networking and Services}.\hskip 1em plus 0.5em minus 0.4em\relax New York, NY, USA: Association for Computing Machinery, 2017, pp. 393--402. [Online]. Available: \url{https://doi.org/10.1145/3144457.3144468}
\BIBentrySTDinterwordspacing

\bibitem{Wang2016}
\BIBentryALTinterwordspacing
W.~Wang, A.~X. Liu, and M.~Shahzad, ``Gait recognition using wifi signals,'' in \emph{Proceedings of the 2016 ACM International Joint Conference on Pervasive and Ubiquitous Computing}, ser. UbiComp '16.\hskip 1em plus 0.5em minus 0.4em\relax New York, NY, USA: Association for Computing Machinery, 2016, pp. 363--373. [Online]. Available: \url{https://doi.org/10.1145/2971648.2971670}
\BIBentrySTDinterwordspacing

\bibitem{Guo2017}
L.~Guo, L.~Wang, J.~Liu, W.~Zhou, B.~L.~T. Liu, G.~Li, and C.~Li, ``A novel benchmark on human activity recognition using wifi signals,'' in \emph{2017 IEEE 19th International Conference on e-Health Networking, Applications and Services (Healthcom)}, 2017, pp. 1--6.

\bibitem{Tan2016}
\BIBentryALTinterwordspacing
S.~Tan and J.~Yang, ``Wifinger: leveraging commodity wifi for fine-grained finger gesture recognition,'' in \emph{Proceedings of the 17th ACM International Symposium on Mobile Ad Hoc Networking and Computing}, 2016, pp. 201--210. [Online]. Available: \url{https://doi.org/10.1145/2942358.2942393}
\BIBentrySTDinterwordspacing

\bibitem{Ma2018}
\BIBentryALTinterwordspacing
Y.~Ma, G.~Zhou, S.~Wang, H.~Zhao, and W.~Jung, ``Signfi: Sign language recognition using wifi,'' \emph{Proc. ACM Interact. Mob. Wearable Ubiquitous Technol.}, vol.~2, no.~1, mar 2018. [Online]. Available: \url{https://doi.org/10.1145/3191755}
\BIBentrySTDinterwordspacing

\bibitem{Tian2018}
Z.~Tian, J.~Wang, X.~Yang, and M.~Zhou, ``Wicatch: A wi-fi based hand gesture recognition system,'' \emph{IEEE Access}, vol.~6, pp. 16\,911--16\,923, 2018.

\bibitem{Ren2019}
S.~Ren, H.~Wang, L.~Gong, C.~Xiang, X.~Wu, and Y.~Du, ``Intelligent contactless gesture recognition using wlan physical layer information,'' \emph{IEEE Access}, vol.~7, pp. 92\,758--92\,767, 2019.

\bibitem{Chen2022}
Y.~Chen, R.~Ou, Z.~Li, and K.~Wu, ``Wiface: Facial expression recognition using wi-fi signals,'' \emph{IEEE Transactions on Mobile Computing}, vol.~21, no.~1, pp. 378--391, 2022.

\bibitem{Zhang2017}
D.~Zhang, H.~Wang, and D.~Wu, ``Toward centimeter-scale human activity sensing with wi-fi signals,'' \emph{Computer Society}, vol.~50, no.~1, pp. 48--57, 2017.

\bibitem{Niu2018}
\BIBentryALTinterwordspacing
K.~Niu, F.~Zhang, Z.~Chang, and D.~Zhang, ``Um sistema de detecção de respiração humana baseado em modelo de difração de fresnel usando dispositivos wi-fi cots,'' in \emph{Proceedings of the 2018 ACM International Joint Conference on Pervasive and Ubiquitous Computing}, ser. UbiComp '18.\hskip 1em plus 0.5em minus 0.4em\relax Association for Computing Machinery, 2018. [Online]. Available: \url{https://doi.org/10.1145/3267305.3267561}
\BIBentrySTDinterwordspacing

\bibitem{Zhang2019}
\BIBentryALTinterwordspacing
F.~Zhang, K.~Niu, J.~Xiong, B.~Jin, T.~Gu, Y.~Jiang, and D.~Zhang, ``Rumo a uma abordagem de detecção baseada em difração no reconhecimento da atividade humana,'' \emph{Proc. ACM Interact. Mob. Wearable Ubiquitous Technol.}, vol.~3, no.~1, março 2019. [Online]. Available: \url{https://doi.org/10.1145/3314420}
\BIBentrySTDinterwordspacing

\bibitem{Kong2021}
\BIBentryALTinterwordspacing
H.~Kong, L.~Lu, J.~Yu, Y.~Chen, X.~Xu, F.~Tang, and Y.-C. Chen, ``Multiauth: Enable multi-user authentication with single commodity wifi device,'' in \emph{Proceedings of the Twenty-Second International Symposium on Theory, Algorithmic Foundations, and Protocol Design for Mobile Networks and Mobile Computing}.\hskip 1em plus 0.5em minus 0.4em\relax New York, NY, USA: Association for Computing Machinery, 2021, pp. 31--40. [Online]. Available: \url{https://doi.org/10.1145/3466772.3467032}
\BIBentrySTDinterwordspacing

\bibitem{Afshar2022}
A.~Afshar, V.~T. Vakili, and S.~Daei, ``Active user detection and channel estimation for spatial-based random access in crowded massive mimo systems via blind super-resolution,'' \emph{IEEE Signal Processing Letters}, vol.~29, pp. 1072--1076, 2022.

\bibitem{Meneghello2023}
F.~Meneghello, D.~Garlisi, N.~D. Fabbro, I.~Tinnirello, and M.~Rossi, ``Sharp: Environment and person independent activity recognition with commodity ieee 802.11 access points,'' \emph{IEEE Transactions on Mobile Computing}, vol.~22, no.~10, pp. 6160--6175, 2023.

\bibitem{Zhao2021}
Y.~Zhao, R.~Gao, S.~Liu, L.~Xie, J.~Wu, H.~Tu, and B.~Chen, ``Device-free secure interaction with hand gestures in wifi-enabled iot environment,'' \emph{IEEE Internet of Things Journal}, vol.~8, no.~7, pp. 5619--5631, 2021.

\bibitem{Cheng2021}
X.~Cheng, B.~Huang, and J.~Zong, ``Device-free human activity recognition based on gmm-hmm using channel state information,'' \emph{IEEE Access}, vol.~9, pp. 76\,592--76\,601, 2021.

\bibitem{Uddin2019}
M.~Uddin, S.~Islam, and A.~Al-Nemrat, ``A dynamic access control model using authorising workflow and task-role-based access control,'' \emph{IEEE Access}, vol.~7, pp. 166\,676--166\,689, 2019.

\bibitem{Yang2023}
Y.-J. Yang, C.-M. Chao, C.-C. Yeh, and C.-Y. Lin, ``Wfid: Driver identity recognition based on wi-fi signals,'' \emph{IEEE Transactions on Vehicular Technology}, vol.~72, no.~1, pp. 679--688, 2023.

\bibitem{PBV2024}
R.~R. PBV, V.~Sonaleo~Mandapati, S.~L. Pilli, P.~Lahari~Manojna, T.~H. Chandana, and V.~Hemalatha, ``Home security with iot and esp32 cam - ai thinker module,'' in \emph{2024 International Conference on Cognitive Robotics and Intelligent Systems (ICC - ROBINS)}, 2024, pp. 710--714.

\bibitem{Lin2018}
C.~Lin, J.~Hu, Y.~Sun, F.~Ma, L.~Wang, and G.~Wu, ``Wiau: An accurate device-free authentication system with resnet,'' in \emph{2018 15th Annual IEEE International Conference on Sensing, Communication, and Networking (SECON)}, 2018, pp. 1--9.

\bibitem{Zou2018}
H.~Zou, Y.~Zhou, J.~Yang, H.~Jiang, L.~Xie, and C.~J. Spanos, ``Deepsense: Device-free human activity recognition via autoencoder long-term recurrent convolutional network,'' in \emph{2018 IEEE International Conference on Communications (ICC)}, 2018, pp. 1--6.

\bibitem{Chen2019}
Z.~Chen, L.~Zhang, C.~Jiang, Z.~Cao, and W.~Cui, ``Wifi csi based passive human activity recognition using attention based blstm,'' \emph{IEEE Transactions on Mobile Computing}, vol.~18, no.~11, pp. 2714--2724, 2019.

\bibitem{Ding2019}
J.~Ding and Y.~Wang, ``Wifi csi-based human activity recognition using deep recurrent neural network,'' \emph{IEEE Access}, vol.~7, pp. 174\,257--174\,269, 2019.

\bibitem{Gu2021}
Y.~Gu, H.~Yan, M.~Dong, M.~Wang, X.~Zhang, Z.~Liu, and F.~Ren, ``Wione: One-shot learning for environment-robust device-free user authentication via commodity wi-fi in man--machine system,'' \emph{IEEE Transactions on Computational Social Systems}, vol.~8, no.~3, pp. 630--642, 2021.

\bibitem{Gu2022}
Y.~Gu, Y.~Wang, M.~Wang, Z.~Pan, Z.~Hu, Z.~Liu, F.~Shi, and M.~Dong, ``Secure user authentication leveraging keystroke dynamics via wi-fi sensing,'' \emph{IEEE Transactions on Industrial Informatics}, vol.~18, no.~4, pp. 2784--2795, 2022.

\bibitem{Mitra2022}
A.~Mitra, D.~Bigioi, S.~P. Mohanty, P.~Corcoran, and E.~Kougianos, ``iface 1.1: A proof-of-concept of a facial authentication based digital id for smart cities,'' \emph{IEEE Access}, vol.~10, pp. 71\,791--71\,804, 2022.

\bibitem{Yousefi2017}
S.~Yousefi, H.~Narui, S.~Dayal, S.~Ermon, and S.~Valaee, ``A survey on behavior recognition using wifi channel state information,'' \emph{IEEE Communications Magazine}, vol.~55, no.~10, pp. 98--104, 2017.

\bibitem{YMa2019}
\BIBentryALTinterwordspacing
Y.~Ma, G.~Zhou, and S.~Wang, ``Wifi sensing with channel state information: A survey,'' \emph{ACM Comput. Surv.}, vol.~52, no.~3, pp. 46:1--46:36, jun 2019. [Online]. Available: \url{https://doi.org/10.1145/3310194}
\BIBentrySTDinterwordspacing

\bibitem{DanWu2017}
D.~Wu, D.~Zhang, C.~Xu, H.~Wang, and X.~Li, ``Device-free wifi human sensing: From pattern-based to model-based approaches,'' \emph{IEEE Communications Magazine}, vol.~55, no.~10, pp. 91--97, 2017.

\bibitem{Gringoli2019}
\BIBentryALTinterwordspacing
F.~Gringoli, M.~Schulz, J.~Link, and M.~Hollick, ``Free your csi: A channel state information extraction platform for modern wi-fi chipsets,'' in \emph{Proceedings of the 13th International Workshop on Wireless Network Testbeds, Experimental Evaluation \& Characterization}.\hskip 1em plus 0.5em minus 0.4em\relax New York, NY, USA: Association for Computing Machinery, 2019, pp. 21--28. [Online]. Available: \url{https://doi.org/10.1145/3349623.3355477}
\BIBentrySTDinterwordspacing

\end{thebibliography}
